\documentclass{elsart}

\usepackage{psfig}

\usepackage{amssymb}

\begin{document}

\begin{frontmatter}



\title{Electromagnetic Dissociation as a Tool for Nuclear Structure 
and Astrophysics}


\author[fzj]{Gerhard Baur\thanksref{emgb}}
\thanks[emgb]{E-mail: g.baur@fz-juelich.de}
\author[uba]{Kai Hencken\thanksref{emkh}}
\thanks[emkh]{E-mail: k.hencken@unibas.ch}
\author[uba]{Dirk Trautmann\thanksref{emdt}}
\thanks[emdt]{E-mail: dirk.trautmann@unibas.ch}
\author[umu]{Stefan Typel\thanksref{emst}}
\thanks[emst]{E-mail: stefan.typel@physik.uni-muenchen.de}
\author[umu]{Hermann H. Wolter\thanksref{emhw}}
\thanks[emhw]{E-mail: hermann.wolter@physik.uni-muenchen.de}

\address[fzj]{Forschungszentrum J\"ulich, D-52425 J\"ulich, Germany}
\address[uba]{Universit\"at Basel, CH-4056 Basel, Switzerland}
\address[umu]{Universit\"at M\"unchen, D-85748 Garching, Germany}

\begin{abstract}
Coulomb dissociation is an especially simple and important reaction
mechanism. Since the perturbation due to the electric field of the
nucleus is exactly known, firm conclusions can be drawn from such
measurements. Electromagnetic matrix elements and astrophysical
S-factors for radiative capture processes can be extracted from
experiments.  We describe the basic theory, new results concerning
higher order effects in the dissociation of neutron halo nuclei, and
briefly review the experimental results obtained up to now.  Some new
applications of Coulomb dissociation for nuclear astrophysics and
nuclear structure physics are discussed.
\end{abstract}

\end{frontmatter}

\section{Introduction}
With increasing beam energy higher lying states of nuclei can be
excited with the Coulomb excitation mechanism. This can lead to
Coulomb dissociation, in addition to Coulomb excitation of particle
bound states, for a review see , e.g., \cite{Ber01}.  Such
investigations are also well suited for secondary (radioactive) beams.
In this review we start with a general discussion of Coulomb
dissociation.  Due to the time-dependent electromagnetic field the
projectile is excited to a bound or continuum state, which can
subsequently decay.  We briefly mention the very large effects of
electromagnetic excitation in relativistic heavy ion collisions.  If
1$^{st}$ order electromagnetic excitation is the dominant effect,
experiments can directly be interpreted in terms of electromagnetic
matrixelements, which also enter e.g.  in radiative capture
cross-sections The question of higher order effects is therefore very
important. We present new results for a simple and realistic model for
Coulomb dissociation of neutron halo nuclei. We show that these
effects are reassuringly small.  After a short review of results
obtained for nuclear structure as well as nuclear astrophysics, we
discuss new possibilities, like the experimental study of two-particle
capture.  We close with conclusions and an outlook.

\section{ General Remarks on Electromagnetic Dissociation}
Coulomb excitation is a very useful tool to determine nuclear
electromagnetic matrixelements.  This is of interest for nuclear
structure and nuclear astrophysics \cite{Bau01,Bau02}.  Multiple
electromagnetic excitation can also be important. We especially
mention two aspects: It is a way to excite new nuclear states, like
the double phonon giant dipole resonance \cite{Bau02}; but it can also
be a correction to the one-photon excitation \cite{Typ01,Typ02,Typ03}.

In the equivalent photon approximation the cross section for an
electromagnetic process is written as
\begin{equation}
 \sigma = \int \frac{d\omega}{\omega} \: n(\omega) \sigma_{\gamma}(\omega)
\end{equation}
where $\sigma_{\gamma}(\omega)$ denotes the appropriate cross section
for the photo-induced process and $n(\omega)$ is the equivalent photon
number. For sufficiently high beam energies it is well approximated by
\begin{equation}
 n(\omega) = \frac{2}{\pi} Z^{2} \alpha \ln \frac{\gamma v}{\omega R}
\end{equation}
where $R$ denotes some cut-off radius. More refined expressions, which
take into account the dependence on multipolarity, beam velocity or
Coulomb-de\-flec\-tion, are available in the literature
\cite{Ber01,Typ02,Win01}.  The theory of electromagnetic excitation is
well developed for nonrelativistic, as well as relativistic projectile
velocities.  In the latter case an analytical result for all
multipolarities was obtained in Ref.~\cite{Win01}. The projectile
motion was treated classically in a straight-line approximation. On
the other hand, in the Glauber theory, the projectile motion can be
treated quantally \cite{Ber01,Typ03}.  This gives rise to
characteristic diffraction effects.  The main effect is due to the
strong absorption at impact parameters less than the sum of the two
nuclear radii.

\section{Electromagnetic Excitation in Relativistic Heavy Ion Collisions} 
Electromagnetic excitation is also used at relativistic heavy ion
accelerators to obtain nuclear structure information. Recent examples
are the nuclear fission studies of radioactive nuclei
\cite{khschmidt00} and photofission of $^{208}$Pb \cite{abreu99}.
Cross-sections for the excitation of the giant dipole resonance
(``Weizs\"acker-Williams process'') at the forthcoming relativistic
heavy ion colliders RHIC and LHC(Pb-Pb) at CERN are huge
\cite{RHIC89,Hen02}, of the order of 100 b for heavy systems (Au-Au or
Pb-Pb).  In colliders, the effect is considered to be mainly a
nuisance, because the excited particles are lost from the beam. On the
other hand, the effect will also be useful as a luminosity monitor by
detecting the neutrons in the forward direction.  Specifically one
will measure the neutrons which will be produced after the decay of
the giant dipole resonance which is excited in each of the
ions(simultaneous excitation).  Since this process has a steeper
impact parameter dependence than the single excitation cross-section,
there is more sensitivity to the cut-off radius and to nuclear
effects.  For details and further Refs., see \cite{Hen02}.

\section{Higher Order Effects and Postacceleration}
Higher order effects can be taken into account in a coupled channels
approach, or by using higher order perturbation theory. The latter
involves a sum over all intermediate states $n$ considered to be
important.  Another approach is to integrate the time-dependent
Schr\"{o}dinger equation directly for a given model Hamiltonian
\cite{Mel99,Esb01,paris99,typwo99}.  If the collision is sudden, one
can neglect the time ordering in the usual perturbation approach. Then
the interaction can be summed to infinite order.  Intermediate states
$n$ do not appear explicitly.

Higher order effects were recently studied in \cite{Tos99}, where
further references also to related work can be found. Since full
Coulomb wave functions in the initial and final channels are used
there, the effects of higher order in $\eta_{\rm coul}=\frac{Z Z_c
e^2}{\hbar v} $ are taken into account to all orders. Expanding this
T-matrixelement for the process $Z+a \rightarrow Z+c+n $ in this
parameter $\eta_{\rm coul}$ one obtains the Born approximation
\begin{equation}
 T =f_{coul} \frac{2 D_0}{\pi^2}
 \left(\frac{1}{q_a^2-(\vec{q}_n+\vec{q}_c)^2}
+\frac{m_c}{(m_n+m_c)(q_c^2-(\vec{q}_n-\vec{q}_a)^2)}\right)
\end{equation}
where the zero-range constant $D_0$ is given by $\frac{\hbar^2}{2\mu}
\sqrt{8 \pi \eta}$.  The parameter $\eta$ is related to the binding
energy $E_0$ by $E_0=\frac{\hbar^2\eta^2}{2\mu}$. The quantitiy
$f_{coul}$ is related to the elastic Coulomb scattering amplitude up
to an irrelevant phase factor. It is given by $f_{coul}=\frac{2
\eta_{coul} q_a} {q_{coul}^2}$ , where the momentum transfer is given
by $\vec{q}_{coul}= \vec{q}_a - \vec{q}_c - \vec{q}_n $.  This
expression is somehow related to the Bethe-Heitler formula for
brems\-strahlung. The Bethe-Heitler formula has two terms, one of
which corresponds to a Coulomb interaction between the electron and
the target followed by the photon emission, and another one, where the
photon is emitted first and then the electron scatters from the
nucleus. Here we have a Coulomb scattering of the incoming particle
followed by breakup $a=(c+n) \rightarrow c+n$ and another term, where
the projectile $a$ breaks up into $c+n$, and subsequently, $c$ is
scattered on the target $Z$. In the case of bremsstrahlung it is well
known \cite{landau} that even for $\eta_{coul} \gg 1 $ one obtains the
Born approximation result as long as the scattering is into a narrow
cone in the forward direction.  This leads one to suspect that higher
order effects are not very large in the case of high energy Coulomb
dissociation, where the fragments are emitted into the forward
direction.

For a small enough Coulomb push $q_{coul}$ the T-matrixelement,
eq.~(3) can be expanded in $q_{coul}/q$ to give in lowest order (using
energy conservation)
\begin{equation}
 T= f_{coul} \frac{2 D_0}{\pi^2} \frac{(m_n)^2 m_c}{(m_n+m_c)^3}
\frac{2 \vec{q} \cdot \vec{q}_{coul}}{(\eta^2+q^2)^2}
\end{equation}
The relative momentum between $n$ and $c$ is given by $\vec{q}=
\frac{m_c\vec{q}_n-m_n\vec{q}_c}{m_n+m_c} $.  This formula is in
remarkable agreement with the usual $1^{st}$ order treatment of
electromagnetic excitation in the semiclassical approximation. In this
approach the T-matrix is proportional to the elastic scattering
amplitude times an excitation probability.  This excitation amplitude
is e.g. given explicitly for the present zero range model in
\cite{Typ01}.
 
We investigate higher order effects in the model of
\cite{Typ01,Typ02,Typ03}.  In a zero range model for the neutron-core
interaction, analytical results were obtained for $1^{st}$ and
$2^{nd}$ order electromagnetic excitation for small values of the
adiabaticity parameter $\xi$.  We are especially interested in
collisions with small impact parameters where higher order effects
tend to be larger than for the very distant ones. In this case, the
adiabaticity parameter $\xi$ is small. For $\xi =0$ (sudden
approximation) we have a closed form solution, where higher order
effects are taken into account to all orders. In eq.~(37) of
\cite{Typ01} the angle integrated breakup probability is given.  We
expand this expression in the strength parameter $y= \frac{2 Z Z_c e^2
m_n}{\hbar v (m_n+m_c) b \eta}$, where $b$ is the impact parameter. We
define $x=\frac{q}{\eta}$ where the wave number q is related to the
energy $E_{\rm rel}$ of the continuum final state by $E_{\rm rel}=
\frac{\hbar^2 q^2}{2\mu}$. In leading order (LO) we obtain
\begin{equation}
 \frac{dP_{LO}}{dq}=\frac{16}{3\pi \eta} y^2 \frac{x^4}{(1+x^2)^4}
\end{equation}
The next to leading order (NLO) expression is proportional to $y^4$
and contains a contribution from the $ 2^{nd}$ order E1 amplitude and
a contribution from the interference of $ 1^{st}$ and $ 3^{rd}$ order.
We find
\begin{equation}
 \frac{dP_{NLO}}{dq}=\frac{16}{3\pi\eta} y^4 
 \frac{x^2 (5-55x^2+28 x^4)}{15 (1+x^2)^6}
\end{equation}
The integration over x and the impact parameter b can also be
performed analytically in good approximation. For details see
\cite{tyba}. We can easily insert the values for the Coulomb
dissociation experiments on $^{11}$Be and $^{19}$C \cite{Nak02,Nak03}
in the present formulae.  We find that the ratio of the NLO
contribution to the LO contribution in the case of Coulomb
dissociation on $^{19}$C \cite{Nak03} is given by $-10\%$. This is to
be compared to the results of \cite{Tos99} where a value of about
$-35\%$ was found. The reason for these differences has to be
investigated further.

Postacceleration is a higher order effect. A semiclassical model might
suggest that the parallel momentum distribution of the core is shifted
towards larger values due to an "extra Coulomb push", see
e.g. \cite{BBK}. However, this turns out to be wrong. In the sudden
approximation, the core-neutron binding is negligible.  Also on its
way towards the target the core alone( and not the bound neutron- core
system) feels the Coulomb interaction. Thus there is no extra Coulomb
push. Corrections due to finite values of $\xi$ were studied in
\cite{Typ01}.  They were found to be a rather delicate quantal
interference effect depending only on the phase shift of the neutron
s-wave.  The effects are quite small and it is worth mentioning that
no postacceleration effects were found in a recent $^{11}$Be Coulomb
dissociation experiment \cite{bush}.

\section{Discussion of some experimental results for nuclear structure and 
astrophysics}
Coulomb dissociation of exotic nuclei is a valuable tool to determine
electromagnetic matrix-elements between the ground state and the
nuclear continuum.  The excitation energy spectrum of the
${}^{10}$Be+n system in the Coulomb dissociation of the one-neutron
halo nucleus ${}^{11}$Be on a Pb target at $72\cdot$A~MeV was measured
\cite{Nak02}.  Low lying E1-strength was found. The Coulomb
dissociation of the extremely neutron-rich nucleus ${}^{19}$C was
recently studied in a similar way \cite{Nak03}.  The neutron
separation energy of ${}^{19}$C could also be determined to be
$530\pm130$~keV.  Quite similarly, the Coulomb dissociation of the
2n-halo nucleus ${}^{11}$Li was studied in various laboratories
\cite{Kob89,Shi95,Zin97}. In an experiment at MSU \cite{Iek01}, the
correlations of the outgoing neutrons were studied. Within the limits
of experimental accuracy, no correlations were found.

In nuclear astrophysics, radiative capture reactions of the type $ b +
c \to a + \gamma $ play a very important role. They can also be
studied in the time-reversed reaction $ \gamma + a \to b + c \: $, at
least in those cases where the nucleus $a$ is in the ground state.  As
a photon beam, we use the equivalent photon spectrum which is provided
in the fast peripheral collision. Reviews, both from an experimental
as well as theoretical point of view have been given \cite{Bau01}, so
we want to concentrate here on a few points.

\begin{figure}[ht]
\begin{center}
\leavevmode\psfig{file=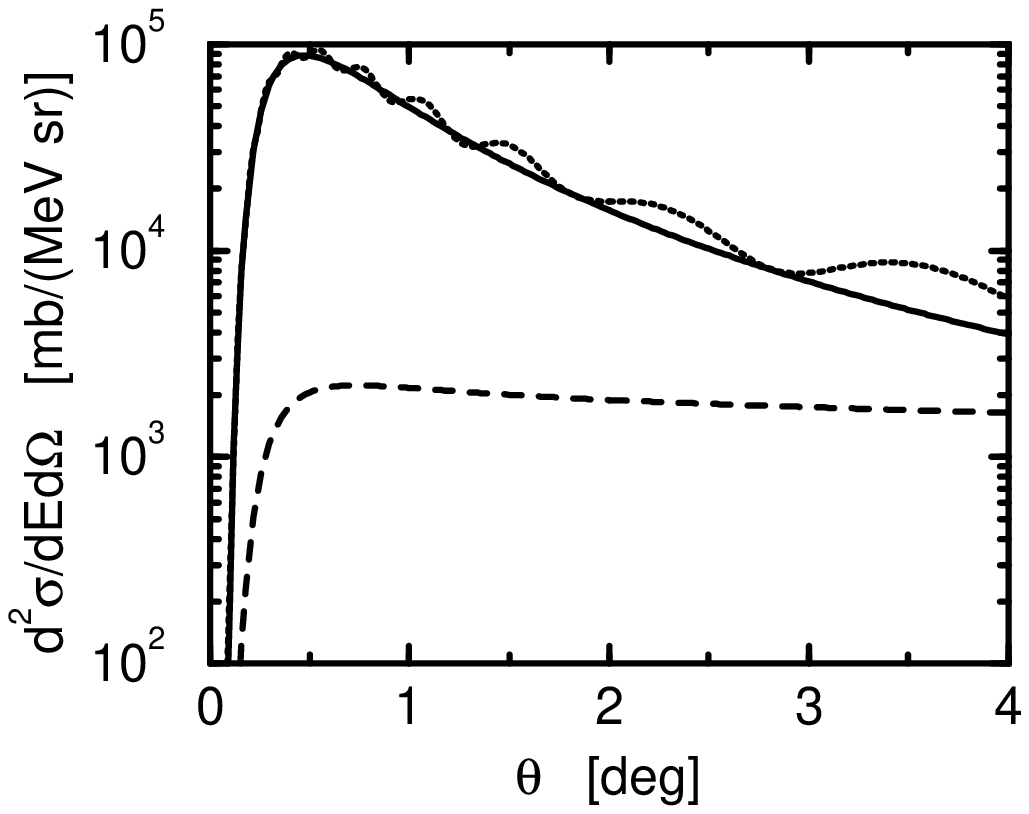,height=5cm,width=5cm}
\leavevmode\psfig{file=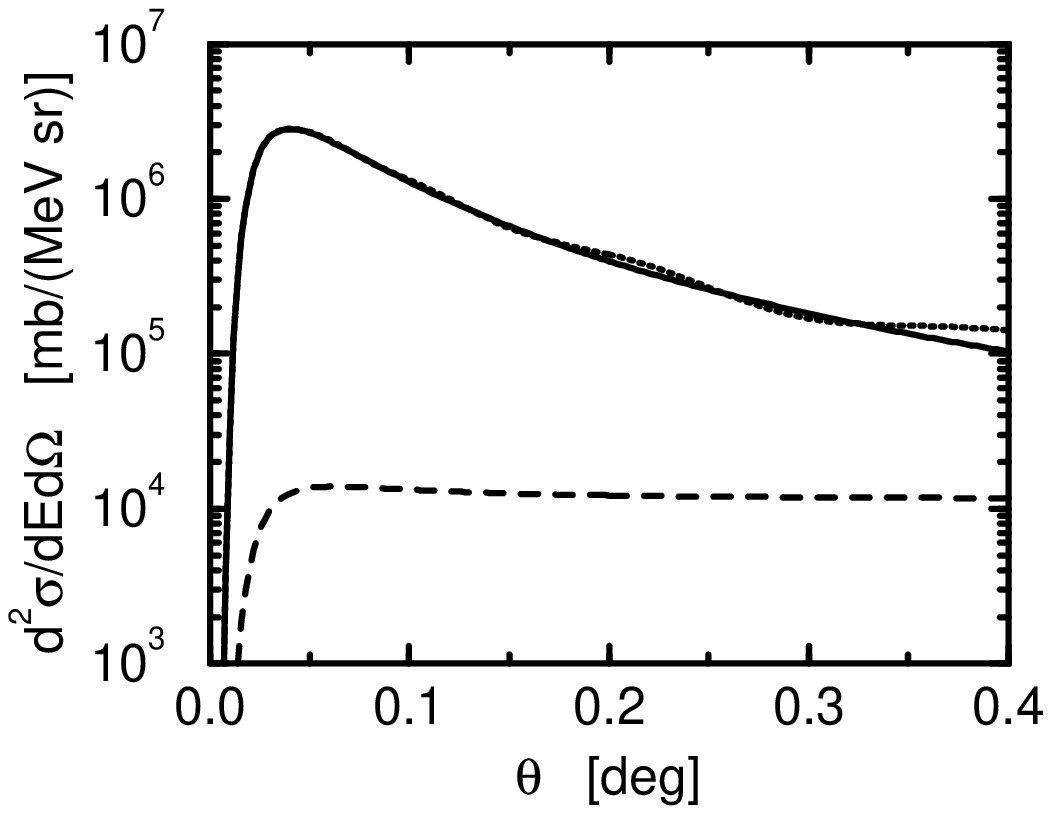,height=5cm,width=5cm}
\end{center}
\caption{
Coulomb dissociation cross section of ${}^{8}$B scattered
on ${}^{208}$Pb as a function of the scattering angle for
projectile energies of 46.5~$A\cdot$MeV (left) and 250~$A\cdot$MeV
(right) and a ${}^{7}$Be-p relative energy of 0.3~MeV. 
First order results E1 (solid line), E2 (dashed line) and
E1+E2 excitation including nuclear diffraction (dotted line). 
[From Figs.~4 and 5 of Ref.~\protect\cite{Typ03}.]
} 
\label{procfig}
\end{figure}

The ${}^{6}$Li Coulomb dissociation into $\alpha$+d has been a test
case of the method, see Ref.~\cite{Bau01}. This is of importance since
the $d(\alpha,\gamma)^6$Li radiative capture is the only process by
which $^6$Li is produced in standard primordial nucleosynthesis
models.  There has been new interest in $^6$Li as a cosmological probe
in recent years, mainly because the sensitivity for searches for
$^6$Li has been improving.  It has been found in metal-poor halo stars
at a level exceeding even optimistic estimates of how much $^6$Li
could have been made in standard big bang nucleosynthesis. For more
discussion on this see \cite{nollett00}.
  
The ${}^{7}$Be(p,$\gamma$)${}^{8}$B radiative capture reaction is
relevant for the solar neutrino problem.  It determines the production
of ${}^{8}$B which leads to the emission of high energy neutrinos.
There are direct reaction measurements, for a recent one see
Refs.~\cite{Ham98}.  Coulomb dissociation of ${}^{8}$B has been
studied at RIKEN \cite{Mot02}, MSU \cite{Kel01} and GSI \cite{Iwa99}.
Theoretical calculations are shown in Fig.~1. It is seen that E1
excitation is large and peaked at very forward angles. E2 excitation
is also present, with a characteristically different angular
distribution. Nuclear diffraction effects are small. Altogether it is
quite remarkable that completely different experimental methods with
possibly different systematic errors lead to results that are quite
consistent.

\section{Possible New Applications of Coulomb dissociation
 for nuclear astrophysics}
Nucleosynthesis beyond the iron peak proceeds mainly by the r- and
s-processes (rapid and slow neutron capture) \cite{Rol01,Cow01}.  To
establish the quantitative details of these processes, accurate
energy-averaged neutron-capture cross sections are needed. Such data
provide information on the mechanism of the neutron-capture process
and time scales, as well as temperatures involved in the process. The
data should also shed light on neutron sources, required neutron
fluxes and possible sites of the processes (see
Ref.~\cite{Rol01}). The dependence of direct neutron capture on
nuclear structure models was investigated in Ref.~\cite{Rau98}.  The
investigated models yield capture cross-sections sometimes differing
by orders of magnitude.  This may also lead to differences in the
predicted astrophysical r-process paths. Because of low level
densities, the compound nucleus model will not be applicable.

With the new radioactive beam facilities (either fragment separator or
ISOL-type facilities) some of the nuclei far off the valley of
stability, which are relevant for the r-process, can be produced. In
order to assess the r-process path, it is important to know the
nuclear properties like $\beta$-decay half-lifes and neutron binding
energies. Sometimes, the waiting point approximation
\cite{Rol01,Cow01} is introduced, which assumes an (n,$\gamma$)- and
($\gamma$,n)-equilibrium in an isotopic chain.  It is generally
believed that the waiting point approximation should be replaced by
dynamic r-process flow calculations, taking into account (n,$\gamma$),
($\gamma$,n) and $\beta$-decay rates as well as time-varying
temperature and neutron density. In slow freeze-out scenarios, the
knowledge of (n,$\gamma$) cross sections is important.

In such a situation, the Coulomb dissociation can be a very useful
tool to obtain information on (n,$\gamma$)-reaction cross sections on
unstable nuclei, where direct measurements cannot be done. Of course,
one cannot and need not study the capture cross section on all the
nuclei involved; rather there will be some key reactions of nuclei
close to magic numbers. It was proposed \cite{Gai01} to use the
Coulomb dissociation method to obtain information about (n,$\gamma$)
reaction cross sections, using nuclei like ${}^{124}$Mo, ${}^{126}$Ru,
${}^{128}$Pd and ${}^{130}$Cd as projectiles. The optimum choice of
beam energy will depend on the actual neutron binding energy. Since
the flux of equivalent photons has essentially an $\frac{1}{\omega}$
dependence, low neutron thresholds are favourable for the Coulomb
dissociation method. Note that only information about the (n,$\gamma$)
capture reaction to the ground state is possible with the Coulomb
dissociation method. The situation is reminiscent of the loosely bound
neutron-rich light nuclei, like ${}^{11}$Be, ${}^{11}$Li and
${}^{19}$C.

A new field of application of the Coulomb dissociation method can be
two nucleon capture reactions. Evidently, they cannot be studied in a
direct way in the laboratory. Sometimes this is not necessary, when
the relevant information about resonances involved can be obtained by
other means (transfer reactions, etc.), like in the triple
$\alpha$-process.

Two-neutron capture reactions in supernovae neutrino bubbles are
studied in Ref.~\cite{Goe01}.  In the case of a high neutron
abundance, a sequence of two-neutron capture reactions,
${}^{4}$He(2n,$\gamma$)${}^{6}$He(2n,$\gamma$)${}^{8}$He can bridge
the $A=5$ and 8 gaps. The ${}^{6}$He and ${}^{8}$He nuclei may be
formed preferentially by two-step resonant processes through their
broad $2^{+}$ first excited states \cite{Goe01}. Dedicated Coulomb
dissociation experiments can be useful, see \cite{au99}. Another key
reaction can be the ${}^{4}$He($\alpha$n,$\gamma$) reaction
\cite{Goe01}.  The ${}^{9}$Be($\gamma$,n) reaction has been studied
directly (see Ref.~\cite{Ajz01}) and the low energy $s_{\frac{1}{2}}$
resonance is clearly established.

In the rp-process, two-proton capture reactions can bridge the waiting
points \cite{Bar01,Goe02,Sch02}. From the
${}^{15}$O(2p,$\gamma$)${}^{17}$Ne,
${}^{18}$Ne(2p,$\gamma$)${}^{20}$Mg and
${}^{38}$Ca(2p,$\gamma$)${}^{40}$Ti reactions considered in
Ref.~\cite{Goe02}, the latter can act as an efficient reaction link at
conditions typical for X-ray bursts on neutron stars.  A ${}^{40}$Ti
$\to$ p + p + ${}^{38}$Ca Coulomb dissociation experiment should be
feasible. The decay with two protons is expected to be sequential
rather than correlated (``${}^{2}$He''-emission).  The relevant
resonances are listed in Table~XII of Ref.~\cite{Goe02}.  In
Ref.~\cite{Sch02} it is found that in X-ray bursts 2p-capture
reactions accelerate the reaction flow into the $Z \geq 36$ region
considerably.  In Table~1 of Ref.~\cite{Sch02} nuclei, on which
2p-capture reactions may occur, are listed; the final nuclei are
${}^{68}$Se, ${}^{72}$Kr, ${}^{76}$Sr, ${}^{80}$Zr, ${}^{84}$Mo,
${}^{88}$Ru, ${}^{92}$Pd and ${}^{96}$Cd (see also Fig.~8 of
Ref.~\cite{Bar01}). It is proposed to study the Coulomb dissociation
of these nuclei in order to obtain more direct insight into the
2p-capture process.

\section{Conclusions}
Peripheral collisions of medium and high energy nuclei (stable or
radioactive) passing each other at distances beyond nuclear contact
and thus dominated by electromagnetic interactions are important tools
of nuclear physics research. The intense source of quasi-real (or
equivalent) photons has opened a wide horizon of related problems and
new experimental possibilities especially for the present and
forthcoming radioactive beam facilities to investigate efficiently
photo-interactions with nuclei (single- and multiphoton excitations
and electromagnetic dissociation).

\section{Acknowledgments}
We have enjoyed collaboration and discussions on the present topics
with very many people. We are especially grateful to C.~A.~Bertulani,
H.~Rebel, F.~R\"{o}sel, and R.~Shyam.

\end{document}